# On the communication between cells of a cellular automaton on the penta- and heptagrids of the hyperbolic plane.

Maurice Margenstern

Laboratoire d'Informatique Théorique et Appliquée, EA 3097,
Université de Metz, I.U.T. de Metz, Département d'Informatique,
Île du Saulcy, 57045 Metz Cedex, France,
*e-mail* : margens@univ-metz.fr

**Abstract**
   This contribution belongs to a combinatorial approach to hyperbolic geometry and it is aimed at possible applications to computer simulations.
   It is based on the splitting method which was introduced by the author and which is reminded in the second section of the paper. Then we sketchily remind the application to the classical case of the pentagrid, *i.e.* the tiling of the hyperbolic plane which is generated by reflections of the regular rectangular pentagon in its sides and, recursively, of its images in their sides. From this application, we derived a system of coordinates to locate the tiles, allowing an implementation of cellular automata.
   At the software level, cells exchange messages thanks to a new representation which improves the speed of contacts between cells. In the new setting, communications are exchanged along actual geodesics and the contribution of the cellular automaton is also linear in the coordinates of the cells.

## INTRODUCTION

Hyperbolic geometry becomes more and more attractive, probably due to the irresistible aesthetic impression given by tilings which are obtained in the hyperbolic plane. An other reason to this attraction seems to be the fact that difficult problems can be solved more easily in the hyperbolic plane. This was first discovered in [22]. But an actual implementation in cellular automata was given in [18,19]. This is the reason why the hyperbolic plane could be seen as a possible tool for application purposes, see [14,2], which are based on the approached defined in [9] and later generalized in [11,12].

First, in section 1, we remind a few features about the hyperbolic plane, in order the paper be self-contained. Then, in the section 2, we present the splitting method, which we introduced in [11,12], with its tightly connected notion of combinatoric tilings. It introduces a general way to define a system of coordinates of the tiles. In section 3, we briefly remind the classical cases of the pentagrid and of the heptagrid. At this occasion, we remind the system of coordinates of the tiles for these tilings. Later, these coordinate systems will be considered as a hardware property of the implementation. In section 4, we remind the results of [17] which allow to go continuously from the pentagrid to the heptagrid and conversely. In section 5, we introduce the new representation at the software level. The new system is closer based on the tree structure of the tiling. We describe a protocol for exchanging messages between cells from which we derive the properties which we announce: the connection between cells through geodesic paths and their computation is in linear or square time with respect to the length of the path.

## 1. THE HYPERBOLIC PLANE

We refer the reader to [24] for introductory material on hyperbolic geometry. In order to fix things and to spare space for this paper, we shall use Poincaré's disc model of the hyperbolic plane, see figure 1, below.

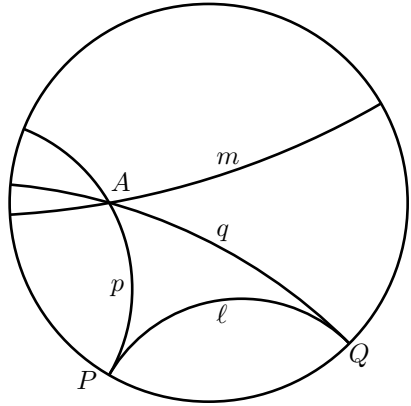

**Figure 1** *The Poincaré's disc as a model of the hyperbolic plane. Here, p and q are parallel to $\ell$ and m does not cut $\ell$.*

The open unit disc $U$ of the Euclidean plane constitutes the points of the hyperbolic plane $I\!H^2$. The border of $U$, $\partial U$, is called the set of **points at infinity**. Lines are the trace in $U$ of its diameters or the trace in $U$ of circles which are orthogonal to $\partial U$. The model has a very remarkable property, which it shares with the half-plane model: hyperbolic angles between lines are the Euclidean angles between the corresponding circles. The model is easily generalised to higher dimension, see [21,16].

This model gives a concrete realisation of hyperbolic geometry within mathematics. It gives an example of a geometry where all the axioms of the Euclidean geometry are satisfied, except the parallel axiom. Here, this axiom is no more true: for any point $A$ out of a line $\ell$, there are exactly two lines which pass through $A$ and which are parallel to $\ell$. There are also infinitely many of them which pass through $A$ and which do not intersect $\ell$, we call such lines **non-secant**. See figure 1, above, where both situations are illustrated.

A direct consequence of this new property about parallel lines is that the sum of the angles of any triangle is less than $\pi$. Indeed, the difference of this sum to $\pi$ is a measure: it is, by definition, the area of the triangle.

We finish with this point by a last consequence: there is no notion of similarity in hyperbolic geometry. In particular, two triangles with the same angles are equal in this geometry, by contrast with what happens in Euclidean geometry.

Recall that in the theory of **tilings**, a particular case consists in considering a polygon $S$ and then, all the reflections of $S$ in its sides and, recursively, of its images in their sides. All the polygons which are obtained in this way are said to be generated by $S$. We say that the set of polygons being generated by $S$ is a **tiling** if and only if the interiors of two distinct polygons either coincide or do not intersect and, if any point of the plane is in the closure of at least one polygon of the set. In this case, we say that the tiling is generated by $S$. A tiling which is obtained in this way is also called generated by **tessellation** from $S$. We shall also say that the tiling is the **grid** defined by $S$.

## 2. THE SPLITTING METHOD
### 2.1. Basis of splitting and spanning tree

**Definition 1** – *Consider finitely many sets $S_0$, ..., $S_k$ of some geometric metric space $X$ which are supposed to be closed with non-empty interior, unbounded and simply connected. Consider also finitely many closed simply connected bounded sets $P_1$, ..., $P_h$ with $h \leq k$. Say that the $S_i$'s and $P_\ell$'s constitute a* **basis of splitting** *if and only if:*

  *(i) $X$ splits into finitely many copies of $S_0$,*

  *(ii) any $S_i$ splits into one copy of some $P_\ell$ and finitely many copies of $S_j$'s,*

*where* **copy** *means an* **isometric image**, *and where, in condition (ii), the copies may be of different $S_j$'s, $S_i$ being possibly included.*

*As usual, it is assumed that the interiors of the copies of $P_\ell$'s and the copies of the $S_j$'s are pairwise disjoint.*

*The set $S_0$ is called the* **head** *of the basis and the $P_\ell$'s are called the* **generating tiles**.

Consider a basis of splitting of $X$, if any. We recursively define a tree $A$ which is associated with the basis as follows. First, we split $S_0$ according to the condition (ii) of definition 1. This gives us a copy of say $P_0$ which we call the **root** of $A$ and which we call also the **leading tile** of $S_0$. In the same way, by the condition (ii) of definition 1, the splitting of each $S_i$ provides us with a copy of some $P_\ell$ which we call the **leading tile** of $S_i$. The splitting provides us also with $k_i$ **regions**, $S_{i_1}$, ..., $S_{i_{k_i}}$ which enter the splitting of $S_i$. The regions which enter the splitting of $S_0$ according to condition (ii) of definition 1 are called the **regions** of the generation 1. Assume that we have all the regions of generation $n$: $S_{n_1}$, ..., $S_{n_{m_n}}$. By definition, their leading tiles constitute the nodes of the generation $n$. We split again these $S_j$'s according to the condition (ii). We obtain $m_n$ tiles which are called the tiles of the generation $n+1$ and, for each $S_{n_h}$ which is some $S_i$, we have a splitting which is the isometric image of the splitting of $S_i$, as above indicated. We say that the leading tiles of these copies of the splitting of $S_i$ are called the **sons** of the leading tile of $S_{n_h}$. By definition, the sons of the leading tile of $S_{n_h}$ belong to the generation $n+1$. The union of all the sons of the nodes of the generation $n$ constitutes the nodes of the generation $n+1$.

This recursive process generates an infinite tree with finite branching. This tree, $A$, is called the **spanning tree of the splitting**, where the *splitting* refers to the basis of splitting $S_0, \ldots, S_k$.

### 2.2. Combinatoric Tilings

We come now to the following general definition:

**Definition 2** − *Say that a tiling of $X$ is* **combinatoric** *if it has a basis of splitting and if the spanning tree of the splitting yields exactly the restriction of the tiling to $S_0$, where $S_0$ is the head of the basis.*

From [11,12], we know that when a tiling is combinatoric, there is a polynomial which is attached to the spanning tree of the splitting. We have the following result:

**Theorem 1** − (Margenstern, [11,12]) *Let $\mathcal{T}$ be a combinatoric tiling, and denote a basis of splitting for $\mathcal{T}$ by $S_0$, …, $S_k$ with $P_0, \ldots, P_h$ as its generating tiles. Let $\mathcal{A}$ be the spanning tree of the splitting. Let $M$ be the square matrix with coefficients $m_{ij}$ such that $m_{ij}$ is the number of copies of $S_{j-1}$ which enter the splitting of $S_{i-1}$ in the condition (ii) of the definition of a basis of splitting. Then the number of nodes of $\mathcal{A}$ belonging to the generation $n$ is given by the sum of the coefficients of the first row of $M^n$. More generally, the number of nodes of the generation $n$ in the tree which is constructed as $\mathcal{A}$ but which is rooted in a node associated to $S_i$ is the sum of the coefficients of the row $i+1$ of $M^n$.*

This matrix is called the **matrix of the splitting** and we call **polynomial of the splitting** the characteristic polynomial of this matrix, possibly divided by the greatest power of $X$ which it contains as a factor. Denote the polynomial by $P$. From $P$, we easily infer the induction equation which allows us to compute very easily the number $u_n$ of nodes of $\mathcal{A}$ on its level $n$. This gives us also the number of nodes of each kind at this level by the coefficients of $M^n$ on the first row: we use the same equation with different initial values. Sequence $\{u_n\}_{n \in \mathbb{N}}$ is called the **recurrent sequence of the splitting**.

Now, as in [9,11,12], number the nodes of $\mathcal{A}$ level by level, starting from the root and, on each level, from the left to the right. Consider the recurrent sequence of the splitting, $\{u_n\}_{n \geq 1}$: it is generated by the polynomial of the splitting. As we shall see, it turns out that the polynomial has a greatest real root $\beta$ with $\beta > 1$. Sequence $\{u_n\}_{n \geq 1}$ is increasing and as well known, see [4,7] for instance, it is possible to represent any positive number $n$ in the form $n = \sum_{i=0}^{k} a_i . u_i$, where $a_i \in \{0..\lfloor \beta \rfloor\}$. The string $a_k \ldots a_0$ is called a representation of $n$. In general, the representation is not unique and it is made unique by an additional condition: we take the representation which is maximal with respect to the lexicographic order on the words on $\{0..b\}$ where $b = \lfloor \beta \rfloor$. The set of these representations is called the **language of the splitting**.

In this paper, we consider only the case when we have a single generating tile, *i.e.* when $h = 1$.

A lot of tilings are combinatoric, see references in [10,13], not only tilings in the hyperbolic plane but also in Euclidean spaces. Moreover, in several cases of tilings of the hyperbolic plane, it turns out that the language of the splitting is regular.

In the next section, we exemplify the application of the method on the pentagrid, the tiling of $I\!H^2$ which is generated by the rectangular regular pentagon.

## 3. THE CLASSICAL CASE OF THE PENTAGRID

Here, the basis of the splitting consists of two regions, $\mathcal{Q}$ and $R_3$. The first region, $\mathcal{Q}$, is a quarter of the hyperbolic plane, the south-western quarter in figure 2. It constitutes the head of the splitting. The second region, $R_3$, which we call a **strip**, appears in figure 2 as the complement in $\mathcal{Q}$ of the union of the leading tile $P_0$ of $\mathcal{Q}$ together with two copies of $\mathcal{Q}$, $R_1$ and $R_2$, see the figure. Region $R_1$ is the image of the shift along side **1** of $P_0$ which transforms a vertex of $P_0$ into the other vertex of the side. Region $R_2$ is obtained from $\mathcal{Q}$ by the shift along side **4** of $P_0$.

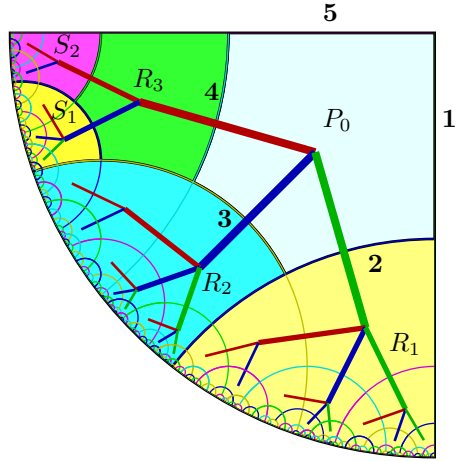

**Figure 2** *The splitting of $\mathcal{Q}$ which is associated to the pentagrid. Notice the construction of the spanning tree.*

This gives us the splitting of $\mathcal{Q}$. We have now to define the splitting of $R_3$.

Its leading tile is provided us by the reflection of $P_0$ in its side **4**, say $P_1$. Call $R'_1$ the image of $\mathcal{Q}$ by the shift along the side **5**. It is also the reflection of $R_1$ in the diagonal of $\mathcal{Q}$. The shift along the side $\mathbf{1}_1$ of $P_1$ transforms $R'_1$ into $S_1$. Now, it is not difficult to see that the complement $S_2$ in $R_3$ of the union of $P_1$ and $S_1$ is the image of $R_3$ under the shift along the side **5** of $P_0$. And so, $R_3$ also can be split according to the definition.

The exact proof that the restriction of the tiling is in bijection with the spanning tree of the splitting is given in [18,20].

In the case of the pentagrid, the number of nodes of the spanning tree which belong to generation $n$ is $f_{2n+1}$ where $\{f_n\}_{n\geq 1}$ is the Fibonacci sequence with $f_1 = 1$ and $f_2 = 2$. This is why the spanning tree of the pentagrid is called the **Fibonacci tree**, starting from [18,20].

From figure 2, and arguing by induction, it is plain that the Fibonacci tree is constructed from the root by the following two rules:

- a 3-node has three sons: to the left, a 2-node and, in the middle and to the right, in both cases, 3-nodes;
- a 2-node has 2 sons: to the left a 2-node, to the right a 3-node.

and starting from the axiom which tells us that the root is a 3-node.

Starting from [9], we call this tree the **standard** Fibonacci tree, because there are a lot of other Fibonacci trees. As it is proved in [9], there are continuously many of them, and each one is associated to a different splitting. We refer the interested reader to [9] or to our previous surveys [10,13].

In [9], a new and more efficient way is defined to locate the cells which lie in the quarter, which is the prefiguration of the algebraic side of the splitting method later introduced in [12]. We number the nodes of the tree with the help of the positive numbers: we attach 1 to the root and then, the following numbers to its sons, going on on each level from the left to the right and from one level to the next one, see figure 3, below.

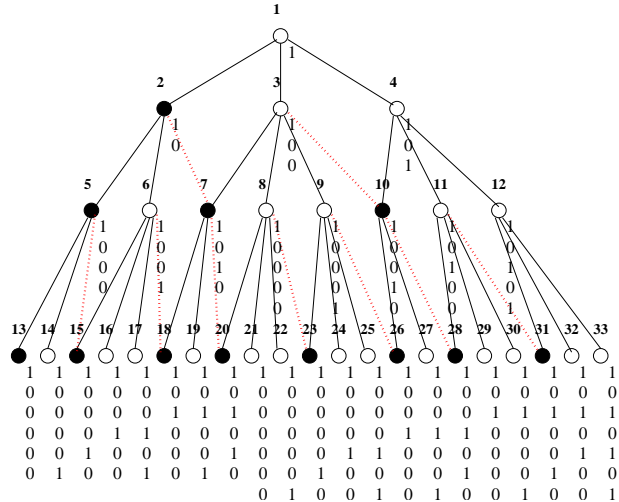

**Figure 3** *The standard Fibonacci tree:*
*above a node: its number; below: its standard representation.*
*Notice that the first node of a level has a Fibonacci number with odd index as its number.*

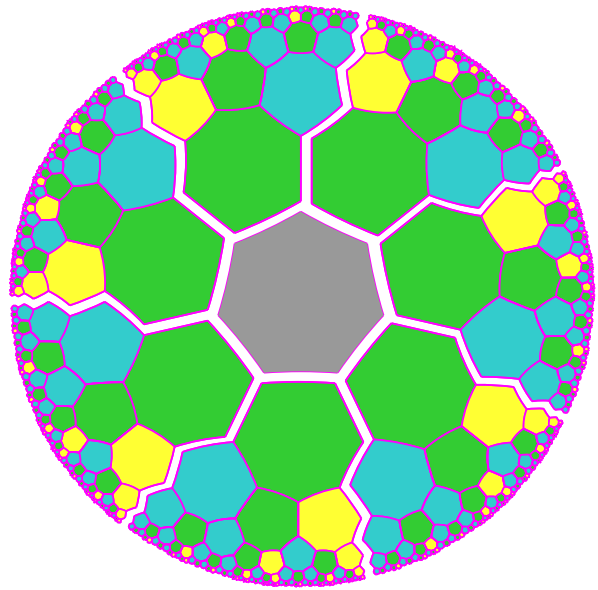

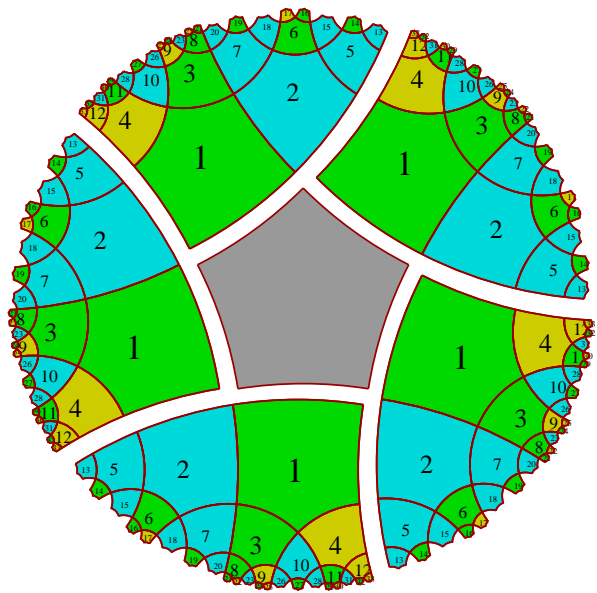

**Figure 4** *Above, the ternary heptagrid by one central tile and seven copies of $S_0$. Below, the pentagrid by one central tile and five copies of $S_0$.*

This numbering is the basis of the system of coordinates which was devised for representing the tiles of the pentagrid starting from [9]. As an example, this system was used in implementations as in [6,3]. The system is the following: as

$I\!H^2$ splits into four copies of $S_0$, it is enough to attach a number in $\{0..3\}$ to each copy and then, inside a copy, to use the just described numbering.

Now, we know that positive integers can be represented with the help of the Fibonacci sequence, again see [7] for instance, which is a very particular case of what was indicated in section 2. Indeed, any positive integer is a sum of distinct Fibonacci numbers. The representation is not unique, in general. It is made unique by an additional condition: we take the longest representation or, equivalently, we do not accept representations which contain the pattern '11'. The longest representation is called the *standard Fibonacci representation*. The language of these representations is regular, which is a classical result in language theory, see [7], and we call **coordinate** of a node the standard Fibonacci representation of its number in $\mathcal{A}$.

## 4. THE FIBONACCI CARPETS

As noted in [17], it is natural to wonder whether it is possible to transport a configuration within the pentagrid into the heptagrid and conversely, without big distortions. In [17], we showed that this is possible and, in the quoted paper, we indicate a solution which is based on the notion of **carpet**.

The idea can be summarized as follows.

In the hyperbolic plane, let us fix an angle $\alpha$ with $\alpha < \pi$. Then the angular sectors of angle $\alpha$ always contain a half-plane. As it is easy to see that $I\!H^2$ is a growing sequence of half-planes, *i.e.* each half-plane is contained in its successor in the sequence, we can expect that $I\!H^2$ has the same property with the angular sectors. And this is the case.

From this property, it is not difficult to conclude that we can do this with quarters: it is the case when $\alpha = \dfrac{\pi}{2}$. As proved in [17], it is not difficult to construct a growing sequence of quarters $\{H_n\}_{n \in \mathbb{Z}}$ possessing the following property. Denoting the Fibonacci tree rooted in the leading tile of $H_n$ by $F_n$, we require that the root of $F_n$ is the middle son of the root of $F_{n+1}$. Of course, we can do the same in the ternary heptagrid. This time there is a growing sequence of angular sectors $\{K_n\}_{n \in \mathbb{Z}}$ which covers $I\!H^2$. These sectors have the angle $\beta$ between the mid-point lines which define a sector in the ternary heptagrid, see [2,17]. Here also, as proved in [17], denoting the Fibonacci tree associated to $K_n$ by $G_n$, we have that the root of $G_n$ is the middle son of the root of $G_{n+1}$.

Now, the bijection between $F_n$ and $G_n$ is easy: it is enough to take tiles with the same coordinate with respect to the corresponding roots.

The **carpet representation** is defined as follows.

Consider a tile $T$ of the pentagrid. There is a unique $n$ such that $T \in F_n$ and $T \notin F_{n-1}$. This $n$ is the first coordinate of $T$. The second coordinate of $T$ is its coordinate $\nu$ in $F_n$. Denoting the coordinates of $T$ by $(n, \nu)$, this define an injection from the tiles of the pentagrid into $\mathbb{Z} \times I\!N$. We can define coordinates in the ternary heptagrid in exactly the same way. The expected bijection consists in putting in correspondence tiles with the same coordinates. As proved in [17], this defines a continuous bijection from the pentagrid into the heptagrid whose inverse is also continuous.

In order to give their full strength to the above results and to enlighten the next section, we define the notion of distance in our context.

A path joining the tile $T_0$ to the tile $T_1$ is a sequence $\{S_i\}_{0 \leq i \leq n}$ of tiles such that:

(i) $S_0 = T_0$, $S_n = T_1$,

(ii) for each $i$, with $0 \leq i < n$, $T_i$ and $T_{i+1}$ have a common side; the number $n$ is called the **length** of the path.

We call **distance** between $T_0$ and $T_1$, denoted by $\mathrm{dist}(T_0, T_1)$, the shortest length for a path joining $T_0$ to $T_1$. It is very easy to show that the distance which we have just defined satisfies the traditional axioms of a metric. A **geodesic** path from $T_0$ to $T_1$ is a path joining $T_0$ to $T_1$ whose length is $\mathrm{dist}(T_0, T_1)$.

We have the following property:

**Lemma 1** *In a Fibonacci tree, there is a single geodesic path from the root to another cell: it is the branch issued from the root which goes through this cell.*

Proof. It is an easy induction on the levels of the tree: the level $n$ is the set of nodes of the tree which are at distance $n$ to the root. ∎

## 5. COMMUNICATIONS BETWEEN THE CELLS

In order to simultaneously deal with both the pentagrid and the heptagrid, we denote by $p$ the number of sides of the polygon which generates the considered tessellation. Accordingly, $p = 5$ for the pentagrid and $p = 7$ for the heptagrid.

In this section, the tiles are called **cells**, as we assume that they have a finite automaton, the same for all the tiles, at their disposal. When we shall consider a cell within a tree and from the point of view of the tree connections, we shall use the term **node** to denote the tile of the cell. As usual for cellular automata, each cell is in contact with its neighbours. Here, the neighbours of a cell $T$ are the cells whose tile shares a side with the tile of $T$.

### 5.1. The problem

Consider the following problem which generalizes the situation considered in [17]. A cell $c$ sends a message to all cells in order to establish communication with the cells which are interested by the message. Such a message will be called **public**. When a cell answers to the message of another which is known, then the message is **private**. As we shall see, the mechanism of routing the messages is not the same depending on whether the conveyed message is public or private. In this paper, we consider only the part of the transmission which is concerned by the computation of the address of the cell to which a message is sent. We do not take into account the size of the message: this raises the question how the message is produced, what are the storage capabilities of the cell and so on. As the resources of a cell of a cellular automaton are bounded, the transmission of messages must be dealt with by the hardware. We shall consider that the cell only indicates whether it has a message or not. In the case of a message, the cell also indicates whether it is public or private. In this context, the only complex thing is the computation of the addresses.

In order to be contacted for a reply, the sender $c$ must send its address to every cell. This will allow a receiver $d$ of the message to send back an acknowledgment or a message accepting the contact to $c$ only. As the address for distant cells may exceed the resource of a cell, the transmission of addresses is also performed by the hardware. However, the cells can contribute to the computation of the address. This is what we assume and we deal with this problem.

In the systems of coordinates of [9], [2] or [17], the addresses being absolute entail that the connection will not be along the shortest path. As proved in [14] and in[17], the path has a length which is linear in the addresses of $c$ and $d$. In the solution which we now suggest in this paper, the path will still be linear in the addresses of $c$ and $d$, but now, it will also be a geodesic path between the cells.

### 5.2. A solution

The idea is to replace absolute coordinates by relative ones and to obtain the address of $d$ in the shortest time.

There is a fixed system of coordinates, for instance, the system described in [2]. The system is based on a central cell $T_0$ with coordinate $(0,0)$ and $p$ angular sectors around $T_0$, covering $I\!H^2$. The $p$ Fibonacci trees associated to the angular sectors together with $T_0$ exactly contain all the tiles of the tessellation.

For each cell $T$ with $T \neq T_0$, its coordinate is $(f,\nu)$ where $f \in \{1..p\}$ is the number of the Fibonacci tree to which it belongs and $\nu$ is the coordinate of $T$ in this tree. This system is used by the hardware in order to locate the cells.

Now, when the cell $c$ emits a public message, $c$ considers itself as the central cell of a system of coordinates. Also, in order to obtain a fast computation, the message arrives to each cell $e$ together with the address of $e$ relative to $c$. Our problem will be to manage the updating of the computed address, each time the message arrives to a new cell. The goal is to realize this operation as fast as possible.

We solve the problem by introducing a new system of coordinates which will be used both as an absolute system and as a relative one. We first describe the system, which we call the **software system** as it is used by the cells during their computations. Then we describe the protocol of the message management and the computations it entails for the cells.

**The software system**

We introduce new **absolute** coordinates for the location of a cell. We describe it in a way which is common for the penta- and the heptagrid. A bit further, we shall indicate another solution which is specific to the pentagrid.

These coordinates consist of digits in $[1..p]$, with $p \in \{5,7\}$, each one representing an **arc** from a node to one of its neighbours. Except for the central cell which has no father, all the other cells have a father, and the arc joining a node to its father has the number 1. The other neighbours of the corresponding cell are joined by the arcs numbered from 2 up to $p$, while counter-clockwise running around the node. This also holds for the roots of the Fibonacci trees which are considered as the sons of the central cell. From this principle, we remark that a given arc does not receive the same number, depending on which sense it is crossed. For instance, an arc 1 goes from a node to its father. For the reverse direction, the number of the same arc is 2, 3 or 4, for the pentagrid, 3, 4 or 5, for the heptagrid, depending on the statuses of the nodes which are at the ends of the arc. In order to avoid problems connected with this difference in the denotation of an arc, we merge both notations in a unique one as follows. Consider an arc from the side 7 of a cell $c$, in the case of the heptagrid. On the other side of this edge, the same side is numbered 2 or 3, depending on the status of the other cell $d$ which shares the side with $c$. And so, if we go from $c$ to $d$, the side will be denoted by $(7,2)$ or $(7,3)$. If we go from $d$ to $c$, the arc will be denoted by $(2,7)$ or $(3,7)$ respectively. We introduce a mirror operation on such couples: let $\delta = (\alpha, \beta)$ with $\alpha, \beta \in \{1..p\}$. We denote the mirror of $\delta$ by $\overline{\delta} = (\beta, \alpha)$. In $\delta = (\alpha, \beta)$, we say that $\alpha$ is the **input** and $\beta$ is the **output**. We define corresponding operators on $\delta$ by: $\alpha = (\delta)_i$ and $\beta = (\delta)_o$. Now, the definition of the mirror will allow us to shorten the notations: if we consider all possible couples of input and output in this order, the set of couples is divided into two equal classes which are the mirror of each other. Accordingly, we introduce the notations of table 1.

For the digit which corresponds to the connection from the central cell to a neighbour or for the opposite connection, we have a similar convention. However, the connections are of the form $(i, 1)$ with $i \in \{1..p\}$. To avoid confusions, we encode these couple with bold digits in $\{\mathbf{1}..\mathbf{p}\}$.

for the heptagrid:

| | |
|---|---|
| $\overline{1} == (1,3)$ | $1 == (3,1)$ |
| $\overline{2} == (1,4)$ | $2 == (4,1)$ |
| $\overline{3} == (1,5)$ | $3 == (5,1)$ |
| $\overline{4} == (2,6)$ | $4 == (6,2)$ |
| $\overline{5} == (2,7)$ | $5 == (7,2)$ |
| $\overline{6} == (3,7)$ | $6 == (7,3)$ |

for the pentagrid:

| | |
|---|---|
| $\overline{1} == (1,2)$ | $1 == (2,1)$ |
| $\overline{2} == (1,3)$ | $2 == (3,1)$ |
| $\overline{3} == (1,4)$ | $3 == (4,1)$ |
| $\overline{4} == (2,5)$ | $4 == (5,2)$ |

**Table 1** *The digits.*

The central cell has 0 as its coordinate.

For the other nodes, the first digit indicates in which Fibonacci tree the node is. We call this first digit the **sector**. Next, the other digits are established according to the indicated principles. The considered sequence can be seen as a variant of the system of digits which occur in the linear algorithm defined in [14].

**The protocol**

The same system is used for the exchange of messages.

When a cell $c$ sends a public message, it considers itself as the centre of a similar system of coordinates which we call **relative** by contrast with the absolute system. The relative address of $c$ is 0. The message is sent to all the other cells of the plane. The other cells behave as relays: they receive the message sent by $c$ together with a sequence of digits $\delta_0..\delta_i$ which is the sequence of arcs going from $c$ to the relay; the relay forwards the message to its sons, within the relative tree rooted in $c$, and transmits them the sequence $\delta_0..\delta_i\delta_{i+1}$. Of course, for each son of the relay, $\delta_{i+1}$ is the arc joining the relay to the son. Accordingly, the work of the relay is only to append $\delta_{i+1}$. But here, we require more: the $\delta_i$'s must denote arcs in the absolute system of coordinates. Now, the computation of $\delta_{i+1}$, as an absolute arc, is easy. Assume that $\delta_i = (\alpha_0, \beta_0)$ and $\delta_{i+1} = (\alpha_1, \beta_1)$. We may assume that each cell knows what is the number of its edges in the other cell of the tiling which shares the same side. This information can be contained in a table *output* which is memorized in the states of the cell. Then, denoting a son by $s$, with $s \in \{0..2\}$, from the leftmost to the rightmost son, $\alpha_1$ and $\beta_1$ are given by:

$$\alpha_1 = 1 + ((\beta_0-1) + 2 + \frac{p-5}{2} + s - st_r) \bmod p, \tag{1}$$
$$\beta_1 = output(\alpha_1).$$

where $st_r$ is the status of the relay in the **relative** tree.

Now, consider the case of a private message. It can always be considered as a reply to a message, either public or private.

If a relay $d$ wishes to reply to the message, it knows its own address from the point of view of $c$. But the elements of the address are absolute arcs. And so, backward taking the same path leads to $c$: the structure of the digits allows to perform this very easily.

Indeed, for the reply, $d$ will ask the hardware to copy the address sequence received from the previous relay which is the address sent by $c$. The cell $d$ knows that the first relay to which it has to send the reply is on the other side of its edge $\beta_i$. And so, as $d$ used the information contained in $\delta_i$ it asks the hardware to prepare two sequences to be transmitted to $c$: $\delta_0..\delta_{i-1}$ and $\delta_i$. The relays will interpret these sequences as stacks whose top is on the right-hand end of the sequence, considered as a word. Accordingly, when a relay receive the reply with the following two sequences: $\delta_0..\delta_j$ and $\delta_i..\delta_{j+1}$, it asks the hardware to pop the top of the first sequence and to push it on the second sequence. And so, it will send the following two sequences $\delta_0..\delta_{j-1}$ and $\delta_j..\delta_i$ to the cell which is on the other side of its edge $(\delta_j)_o$.

It is easy to see that possible further exchanges between $c$ and $d$ will use the same mechanism. Both the strings will be sent with exchanging the rôles each time the receiver is reached. In the direction from $c$ to $d$, the relays read $(\delta)_i$, where $\delta_i$ is the top of the first sequence, while in the direction from $d$ to $c$, the relays read $(\delta)_o$, where $\delta_i$ is also the top of the first sequence. Of course, there is a bit also sent with the sequences which indicates which part of the information should be read.

**An example**

First, consider the example in the heptagrid, see figure 5:

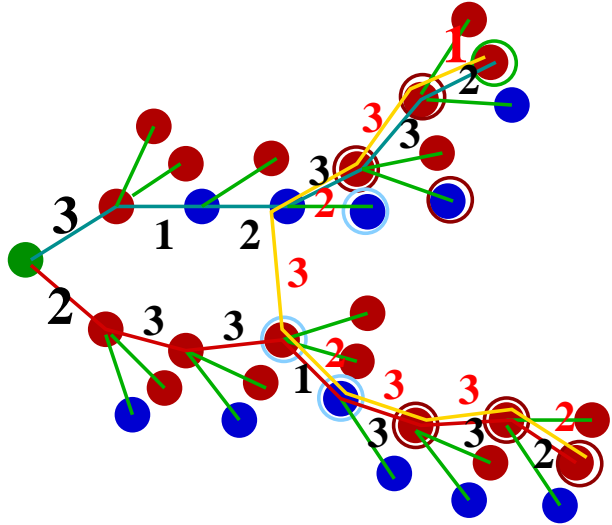

**Figure 5** *Illustration of the example in the heptagrid. The digits in light hue are in relative coordinates.*

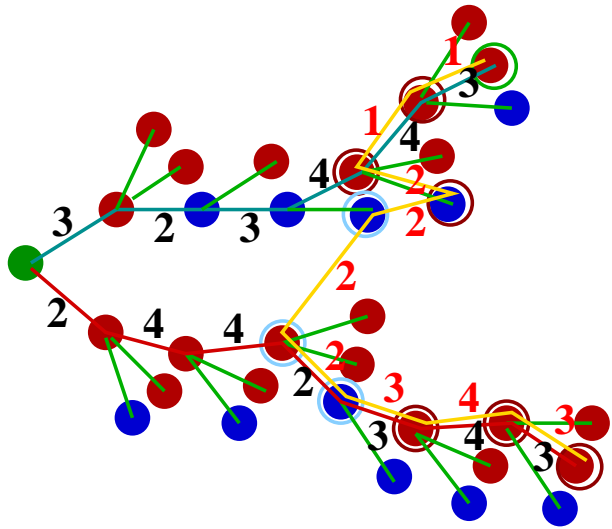

**Figure 6** *Illustration of the example in the pentagrid.*

The sender is the cell with coordinate **3**12332. Consider the cell **2**331332. It is not difficult to see that the address of $d$ with respect to $c$ is $\overline{2336}2332$, see figure 5, while it is **1**3232332 in relative coordinates with $c$ as the central cell. The reverse path is $\overline{2332}6332$ in absolute arcs: it is the reverse of the path from $c$ to $d$.

Note, that if we take the same example in the pentagrid, the coordinates are a bit changed. The absolute coordinate of $c$ is **3**12332, the absolute coordinate of $d$ is **2**331332, but now, the relative coordinate of $d$ from $c$ is **1**3232332, and the same path in absolute coordinates is $\overline{3323}2332$.

Note that when a cell is inside a tree defined by the central cell, the digits of its coordinates are in $\{1..3\}$. The occurrence of other coordinates and especially of over-lined digits indicate either the crossing to another tree or sub-tree or a backward segment of the path.

On the example of figure 6, note that the path from $c$ to $d$ is not exactly the same if we compare this path when in the pentagrid, when in the heptagrid. The crossing from one tree to the neighbouring one does not occur in the same way. This can be explained by the fact that in the heptagrid, each cell has more neighbours than in the case of the pentagrid and, sometimes, this gives a shorter path as we can see by comparing the figures 5 and 6. If we consider the central cell of these figures as an ordinary node, we can see that the phenomenon indicated by the figures also occurs inside trees.

**A special system for the pentagrid**

In the pentagrid, due to the interior angle of $\dfrac{\pi}{2}$, we can use the following property, first noted in [20] and also more especially studied in [15]. We can define a tiling *à la Wang* in the pentagrid by giving numbers in $\{1..5\}$ to the edges of the pentagons by applying the following process.

We start from a given pentagon on which we fix the numbering: we choose at random one of its edges which receives the number 1. Then, we number the other sides by increasing numbers when running around the pentagon, counter-clockwise. Then this numbering fixes the numbers of all other pentagons in such a way that all edges are always given the same number in the pentagons to which they belong. As this property is not true for the ternary heptagrid due to the angle $\dfrac{2\pi}{3}$, the construction works for the pentagrid only.

We define the absolute and relative coordinates as previously with this difference that the digits are now in $\{1..5\}$, as edges have the same number both from the input and the output sides. However, the algorithm to compute the digits which are appended is a bit different. We have to take into account that the numbering has the following property: if in a pentagon $c$ the number are increasing when run counter-clockwise, they are decreasing for all the neighbours of $c$ when the running is still performed counter-clockwise. Accordingly, there is a notion of **orientation**. By definition it is positive for the central cell and it is changed in the opposite orientation each time an edge is crossed. In this new setting, we assume that the cells know their orientation. In this case, if $\delta_0$ is the top of the sequence, if $st_r$ is the status of the present node in the relative tree and if $or \in \{-1,+1\}$ is the absolute orientation of the cell, the new digit $\delta_1$

corresponding to the son $s$ in the relative tree, with $s \in \{0..2\}$, is given by the following formula:

$$\delta_1 = 1 + ((\delta_0-1) + or\times(2+s-st_r))\mod 5. \qquad (2)$$

If we consider the example of figure 6, the cell $c$ has the coordinate 324142, the cell $d$ has the coordinate 2421413. The path from $c$ to $d$ in absolute coordinate is 242131413.

**Theorem 2** *The algorithm of section 5 using table 1 and formulas 1 allows to establish geodesic communications between any cells of the pentagrid or the ternary heptagrid. The similar algorithm based on the natural colouring of the pentagrid and formula 2 does the same for the pentagrid.*

Proof. The 'geodesic' part of the proof is a direct consequence of lemma 1. It comes from the fact that the address which arrives to any cell contacted by $c$ is built along a branch of a Fibonacci tree. ■

### 5.3 Complexity issues

Now, we have to look at the complexity part of the problem. First, we consider the software aspect: the complexity of the work of the cellular automaton. Then we look at the hardware side of the issue.

**Software side**

We have to look exactly on the conditions of the working of the messages. We remember that the rôle of the cellular automaton consists only on computing a new digit while forwarding a public message.

Accordingly, each cell of the cellular automaton indicates to which neighbour a received message has to be forwarded. Each cell also has to compute the part of the transmitted address which has to be appended to the computed address in the case of a public message. In the case of a private message, the mechanism of popping a stack and pushing the popped information on the other stack can be handled by the hardware. This mechanism detaches the popped digit which is transmitted to the cell together with the bit which indicates the direction of reading. The correct input is obtained which allows the cell to get the right output. The table involved in the computation of formula 1 can be stored in the states of the automaton.

In this case, the computation performed by each cell is constant at each top of the clock and so, the above algorithm is linear in time with respect to the distance between the two cells:

**Theorem 3** *The execution of the algorithm considered in theorem 2 is linear in time in the distance between the two communicating cells.*

We also note that at each top of the clock, each cell receives at most $p$ messages as it has $p$ neighbours. As $p$ is a constant of the automaton, it may be contained in the states of the cells. Accordingly, we may assume that the computation of the new digit for the address of a public message is simultaneously performed for all the $p$ messages received by the cell between two consecutive tops of the clock.

However, we have to take into account that a cell may receive several messages for the same destination. It is not difficult to consider cases when this number exceeds the resource of a cell: imagine that all the cells at distance $k$ reply to $c$. If all messages to the same cell are freely transmitted, there are $pf_{2k+1}$ messages arriving at $c$, where $f_n$ is the $n^{\text{th}}$ term in the Fibonacci sequences defined by $f_0 = f_1 = 1$. For this reason, we require that when more than one message arrive at a cell $d$ for the same output, then only one message is transmitted. The others are stored at $d$ in an appropriate stack provided by the hardware. At each top of the clock, the cell gives priority to the messages waiting in the stack which works in a *FIFO* mode.

In this way, messages are guaranteed to reach their goal and there cannot be dead-locks in the system. This includes the possibility for any cell cell relaying private or public messages may also decide to send a public message. But, there may be delays, possibly important, during the transmission. However, if we assume that communications with distant cells are rare, the above situation will not occur and delays will also be rare.

Indeed, assume that the probability for two cells $c$ and $d$ to start a communication is $Ce^{-\lambda \text{dist}(c,d)}$, where $C$ and $\lambda$ are appropriate constants. Let $P_k$ be the probability for $c$ to have a communication with a cell at distance $k$. Then, as the number of cells at distance $k$ from $c$ is $pf_{2k+1}$, we have that $P_k = Cpf_{2k+1}e^{-\lambda \text{dist}(c,d)}$. Now, we know that $f_{2k+1} \sim C_1(\beta)^k$, where $\beta$ is the dominant root of $X^2 - 3X + 1$ and $C_1$ is an appropriate constant. A simple computation gives us that $\beta = \dfrac{3+\sqrt{5}}{2} < e$. Accordingly, $P_k \leq D\gamma^k$, with $D = CC_1p$ and $\gamma = \dfrac{\beta}{e^\lambda}$. If we denote by $P_{\leq k}$ and $P_{>k}$ the probabilities for $c$ to have a communication with a cell at a distance at most $k$ or greater than $k$ respectively, we have that $P_{>k} \leq D\dfrac{\gamma^{k+1}}{1-\gamma}$ and, consequently, $P_{\leq k} > 1 - D\dfrac{\gamma^{k+1}}{1-\gamma}$. We fix $D$ and $\lambda$ so that $\sum_{k=0}^{\infty} P_k = 1$. This computation shows us that our hypothesis of rare distant communications is sound. Accordingly, we may consider that the size of the delays and their occurrence as bounded.

At last, another important aspect of the considered system is that the same message sent by one cell arrives at a given cell exactly once. This comes from the bijection between the tiling and the union of $p$ Fibonacci trees together with the central cell.

**Hardware side**

At the hardware level, we note that going from one cell to another involves the computation of the new absolute address thanks to the given absolute ad-dress. With the system of coordinates which is indicated at the beginning of section 5.2, taken from [9,14], the just considered computation is linear with respect to the given coordinate. Now, from going to one end of a path to the other, the computation is then quadratic as the length of the path is linear in coordinates of the ends of the path.

Now, consider the mechanisms which we introduced in section 5.2. We noted that the management of the stacks involved in these mechanisms must be performed by the hardware. As the total length of the stack is always equal to the length of the path, for each relay, the management of the stacks is linear in the length of the path. And so, the cost of this management is quadratic in the length of the path. And so, at the hardware level, the new system has exactly the same complexity as the absolute system based on the coordinates defined in [9]. As we may assume a bigger speed for the hardware, the quadratic part of the hardware manipulation may be not higher than the linear cost at the level of the cellular automaton for distances which are not higher than the ratio between software and hardware speeds.

For what is messages, the transmission from a cell $c$ to a neighbour consists in copying the message present in $c$ to the appropriate support of $d$. If $M$ is the length of the message, the cost of the copy is $M\rho$, where $\rho$ is the time cost of an elementary hardware operation. Accordingly, if $M$ is the length of a message, the complexity of the transmission from a cell $c$ to a cell $d$ is $M\rho \times \text{dist}(c,d)$. And so, the estimate of the total cost of conveying a message depends on the comparison between $M$ and the distance run by the message. If the distance is big, then the computation of the addresses is more expensive than the pure transmission of the message which is copied from one cell to the next relay. If the length of the message is bigger than the address, then this length is the main factor in the complexity estimation. In fact, the complexity is given by the following result:

**Theorem 4** *The overall complexity of transmitting a message of length $M$ without delay from a cell $c$ to a cell $d$ is $\text{dist}(c,d)\bigl(1+\rho M+\rho\text{dist}(c,d)\bigr)$, where $\rho$ is the time cost of an elementary hardware operation, the time cost of an elementary software operation being $1$.*

## CONCLUSION

As a conclusion of this paper, we hope that the results of this paper can be of help for implement purposes.

It is time to refer here to the few applications of the results quoted from [9,1,2]. These works gave way to one application in human-machine interface, see [3], where a colour palette is presented to allow a user to find easily hues starting from fundamental colours.

We think that there are possible other applications. As indicated in [14], one of them could be the representation of the *Internet*. Another possible application, also mentioned in [14] is special relativity theory. Section 5 could be of interest in both directions.